\documentclass[11pt,twoside]{article}
\usepackage{./asp2014}
\usepackage{rotating}

\aspSuppressVolSlug
\resetcounters

\begin{document}

\title{A Formaldehyde Deep Field}
\author{Jeremy Darling$^1$}
\affil{$^1$Center for Astrophysics and Space Astronomy, Department of Astrophysical and Planetary Sciences,
                 University of Colorado, 389 UCB, Boulder, CO 80309-0389, USA}

\paperauthor{Jeremy Darling}{jeremy.darling@colorado.edu}{0000-0003-2511-2060}{University of Colorado}{Center for Astrophysics and Space Astronomy, Department of Astrophysical and Planetary Sciences}{Boulder}{CO}{80309-0389}{USA}

\begin{abstract}
Formaldehyde (H$_2$CO) is often observed at centimeter wavelengths as an absorption line against the cosmic microwave background (CMB). 
This is possible when energy level populations are anti-inverted to the point where line 
excitation temperatures fall below the local CMB temperature.  Collisions with molecular hydrogen ``pump'' this 
anti-maser excitation, and the cm line ratios of H$_2$CO provide a measurement of the local H$_2$ density.  H$_2$CO 
absorption of CMB light provides all of the benefits of absorption lines (no distance dimming) but none of the drawbacks:  the CMB
provides uniform illumination of all molecular gas in galaxies (no pencil beam sampling), and all galaxies lie in front of the CMB (no fortuitous 
alignments with background light sources are needed).   A formaldehyde deep field (FDF) would therefore provide a blind, mass-limited
survey of molecular gas across the history of star formation and galaxy evolution.  
Moreover, the combination of column density and number density 
measurements may provide geometric distances in large galaxy samples and at higher redshifts than can be done
using the Sunyaev Zel'dovich effect in galaxy clusters.  We present a possible ngVLA FDF that would span redshifts $z=0$--7 and
provide H$_2$CO line ratios over $z=1.36$--3.14.  
\end{abstract}

\section{Background}

The formaldehyde molecule (H$_2$CO) has peculiar non-thermal excitation properties in the physical conditions typical of 
star-forming regions.
Similar to molecules that can be induced to form masers via population inversion through a pumping mechanism, 
H$_2$CO is often ``pumped''  into an anti-inverted state by collisions with molecular hydrogen (H$_2$).  
Anti-inversion is an over-population of a lower-energy state compared to thermal, forming a kind of ultra-cold 
anti-maser that can absorb cosmic microwave background (CMB) photons if the line excitation temperature drops below
the local CMB temperature.
\citet{townes1997} glibly called this effect the ``dasar'' ---  darkness amplification by stimulated absorption of radiation\footnote{Note that darkness is not in fact amplified, nor is absorption stimulated.}.   Unlike maser excitation, which requires special local conditions, 
this anti-inversion of H$_2$CO is nearly ubiquitous in the Galaxy and in external galaxies \citep[e.g.][]{ginsburg2011,mangum2013} 
and seems to be the natural state of the molecule for a wide range of physical conditions \citep[e.g.,][]{darling2012}.  
Moreover, H$_2$CO is observed wherever CO is present and is not strictly a high density molecular gas tracer.  The anti-inverted
$K$-doublet lines are not optically thick and their ratio is a measure of the H$_2$ number density (via collisional pumping).  
The dasar effect can thus be used to make a census of molecular gas mass and gas density, 
independent of redshift (as described below).

The goal of this work is to examine the feasibility of and science enabled by an ngVLA formaldehyde deep field (FDF).  
We assume a flat cosmology with $H_0 = 70$~km~s$^{-1}$~Mpc$^{-1}$, $\Omega_m = 0.3$, and $\Omega_\Lambda = 0.7$.

\subsection{Formaldehyde Anti-Inversion}\label{subsec:anti-inversion}
\begin{figure}
\centering
\includegraphics[width=0.6\textwidth,trim=0 150 70 90,clip]{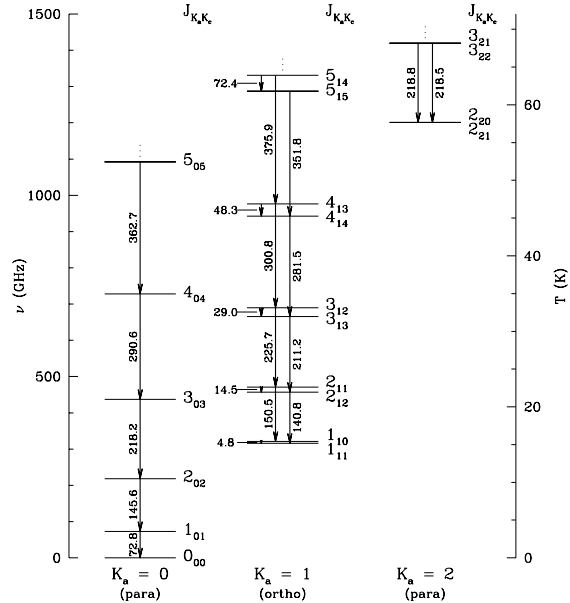}
\caption{Formaldehyde energy levels for the first three
rotation ladders and $J\leq5$ \citep[after][]{darling2012}.  Allowed transition frequencies are listed in GHz.
The anti-inverted cm lines are the $K$-doublet transitions in the $K_a=1$ ortho-H$_2$CO rotation ladder at 4.8, 14.5, 
29.0, 48.3, and 72.4 GHz (6, 2, 1, 0.6, and 0.4~cm, respectively).
\label{H2CO_levels}}
\end{figure}

H$_2$CO is an asymmetric top molecule, which creates a splitting of the primary rotation states (denoted by quantum number
$J$; see Figure \ref{H2CO_levels}).  H$_2$ collisions can overpopulate
the lower energy states of these $K$-doublet rotation states, creating excitation temperatures below the local cosmic microwave background (CMB) temperature by roughly 1--2~K.  The observational signature of this anti-inversion is absorption against the CMB in the centimeter wavelength $K_a = 1$ ortho-H$_2$CO lines primarily at 1, 2, and 6 cm.

This effect is strongest for the lowest energy cm lines and is insensitive to the local gas kinetic temperature.  
The observed line excitation temperature
is also insensitive to the local CMB temperature (which scales from its value of $T_0 = 2.73$~K today as $T_{\rm CMB}(z) =  (1+z)\,T_0$).  The anti-inversion 
favors a fairly wide range of H$_2$ density, roughly $10^2$--$10^5$~cm$^{-3}$, and the cm line ratios indicate the local gas 
number density.  See \citet{darling2012} for detailed modeling versus redshift,
\citet{mangum2008} and \citet{mangum2013} for physical conditions derived 
from observations of H$_2$CO in nearby star-forming galaxies, and \citet{zeiger2010} for a study of H$_2$CO anti-inversion 
in the gravitational lens B0218+357 at $z=0.68$.  

\subsection{Absorption of CMB Light}
The absorption of CMB photons by H$_2$CO implies that the line strength in beam-matched observations is independent of distance.  
And for $z \gtrsim 1$, beam-matching is no longer a strong function of distance because angular sizes become flat for $z\simeq 1$--3
and grow thereafter (Figure \ref{fig:angsize}).  The unusual circumstances presented by H$_2$CO anti-inversion have 
some compelling consequences:  

\smallskip
\noindent
(1) Absorption lines do not require fortuitous alignment of the object of interest with an illuminating light source.  The CMB lies behind every
galaxy and therefore every galaxy with molecular gas may be studied in H$_2$CO absorption.  

\smallskip 
\noindent 
 (2) Unlike traditional absorption line studies, the illuminating ``beam'' is not a pencil beam that samples a subset of the intervening 
galaxy or gas cloud.  The CMB provides an illuminating screen that is uniform to parts in $10^5$ in the CMB rest frame.  All gas is 
sampled in a manner similar to emission line observations (but absorption does not diminish with distance).  

\smallskip 
\noindent 
(3) The consequence of the above two points and the distance-independent nature of absorption lines is that it is possible 
to survey all H$_2$CO gas in the universe in a mass-limited fashion, provided one can beam-match to molecular gas regions in 
galaxies while achieving sub-Kelvin surface brightness sensitivity.

\smallskip 
\noindent
(4) H$_2$CO absorption lines provide the column density of gas, and H$_2$CO line ratios provide the local gas number density.  This 
implies that Sunyaev-Zel'dovich-like distance measurements are possible using the molecular gas in galaxies \citep{darling2012}.  
In contrast to S-Z measurements of the X-ray gas in clusters that extend to $z\sim1$, the H$_2$CO geometric distance measurement 
may be possible in galaxies up to $z\sim6$.  

\smallskip
\noindent
Given these consequences of the peculiarities of the H$_2$CO molecule, one can therefore consider a formaldehyde ``deep field'' for the ngVLA:  a blind, gas mass-limited survey of star-forming galaxies across the history of cosmic star formation.

\section{Angular Resolution and Redshift Coverage}

\begin{figure}[t!]
\plotone{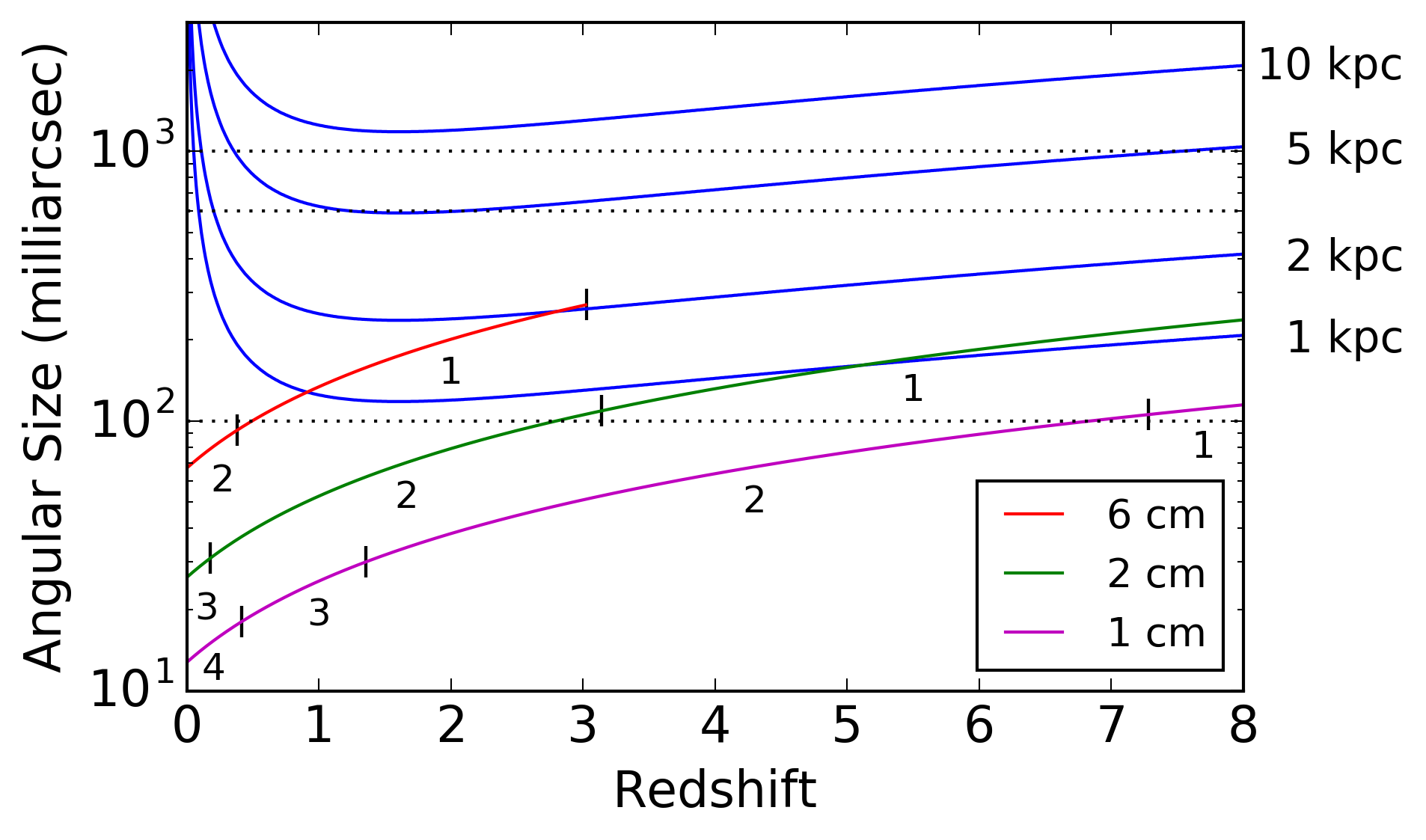}
\caption{
Angular size of various linear size scales (blue lines) and ngVLA beams versus redshift (red, green, and magenta for 
the 6, 2, and 1 cm lines, respectively).  Three H$_2$CO lines are included, and 
the ngVLA bands are labeled by number.  When a ngVLA beam lies below a galaxy size scale (of a disk or starburst nucleus, 
for example), then it is possible to resolve or beam-match that scale.  If most cases, the ngVLA will not suffer beam dilution 
when observing star-forming galaxies or starburst nuclei.  Black dotted lines indicate the fiducial 1 arcsec, 600 mas, and 100 mas
resolutions used for surface brightness calculations (Section \ref{sec:sensitivity}).}\label{fig:angsize}
\end{figure}

In order to avoid beam dilution against the CMB light screen, observations should be beam-matched to the size of star-forming regions in 
galaxies. 
Figure \ref{fig:angsize}  shows angular size tracks versus redshift for 10, 5, 2, and 1 kpc scales.  It also shows the angular
resolution of the ngVLA versus redshift for the 6, 2, and 1 cm H$_2$CO lines and the redshift range covered by each ngVLA band for 
each line.  When the beam size for a given band and redshift is below a given angular size track then that physical scale is resolved.  
Remarkably, the ngVLA can resolve or roughly beam-match 1 kpc at all redshifts in the 1 and 2 cm lines.  Scales of 2 kpc or greater
can be resolved in the 6 cm line up to $z=3$.  The ngVLA can therefore beam-match the physical scales relevant to star-forming 
galaxies including compact starbursts or major mergers (but see Section \ref{sec:sensitivity} for a discussion 
of brightness temperature sensitivity).  Here we examine possible band choices for a FDF:

\subsection{Band 1}\label{sec:band1}

The ngVLA Band 1 (1.2--3.5 GHz) spans $z=0.38$--3.02 in the 6 cm line and $z=3.14$--11.07 in the 2 cm line  (Figure \ref{fig:angsize}). 
In one observation 
covering the full 2.3 GHz bandwidth, it would be possible to nearly span the history of star formation, from 9.3 Gyr
to only 0.4 Gyr after the Big Bang (although molecular gas is unlikely to be detected earlier than $z\sim7$, or $\sim$0.7 Gyr).  
Band 1 would be able to beam-match $\geq 2$ kpc scales in the 6 cm line and $\geq 1$ kpc scales in the 2 cm line spanning the 
above redshift ranges.  Band 1 alone would not provide any H$_2$CO line ratios, so a FDF using only Band 1 would provide a
census of star formation and molecular gas mass (modulo abundance variations between galaxies), but it would not provide
H$_2$ densities.

\subsection{Band 2}\label{sec:band2}

Band 2 (3.5--12.3~GHz) spans $z \leq 0.38$ in the 6 cm line, $z=0.18$--3.14 in the 2 cm line, and $z=1.36$--7.28 in the 1 cm line.
The redshift overlap of the 2 cm and 1 cm lines, spanning $z=1.36$--3.14, would enable the multi-line diagnostics described above (Section \ref{subsec:anti-inversion}) near the peak of cosmic star formation history.  There is also a small low-redshift shell, $z=0.18$--0.38, 
where the 6:2 cm line ratio will be available.  Band 2 observations would resolve sub-kpc physical scales at all redshifts 
(Figure \ref{fig:angsize}).

\subsection{Bands 1 and 2}\label{sec:bands12}

The strongest anti-inverted H$_2$CO line is likely to be the 6 cm line, but Band 2 only captures this line at low redshift.  
If Bands 1 and 2 were both observed as part of a FDF, the 6:2 cm line ratio would be available for $z=0.18$--3.02,
the 2:1 cm line ratio for $z>1.36$ (to redshifts where no molecular gas is expected to exist), and 
the 6:1 cm line ratio for $z=1.36$--3.02.  It would therefore be possible to measure $n({\rm H}_2)$ for nearly all 
redshifts, $z > 0.18$, spanning the history of cosmic star formation and galaxy evolution.  The combined Band 1 and Band 2 
FDF would be able to beam-match or resolve $\geq2$~kpc scales in all cases.  The Band 2 field of view would limit the overlap
between the two bands (24.4 arcmin versus 7.3 arcmin FWHM at the frequency centers of Bands 1 and 2, respectively).

\section{Sensitivity}\label{sec:sensitivity}

The observed line temperature depends on the difference between the line excitation temperature and the background 
continuum brightness temperature, redshifted to the observer's reference frame, and on the optical depth.  
For the H$_2$CO $K$-doublet lines, the continuum is the CMB
(but may also include the host galaxy continuum) at the host galaxy's redshift:
\begin{equation}
 \Delta T_{\rm Obs}={T_{\rm x}(z)-T_{\rm CMB}(z)\over1+z}(1-e^{-\tau}).
\label{eqn:DTobs}
\end{equation}
\citet{darling2012} showed that the observed temperature decrement $\Delta T_{\rm Obs}$ is insensitive to redshift or
the local gas temperature and spans a large range in local gas density, 
$10^2$~cm$^{-3}\lesssim n({\rm H}_2) \lesssim 10^5$~cm$^{-3}$.  At low density, the temperature decrement 
approaches zero (the line excitation temperature equilibrates with the CMB), and at high density, the line excitation 
temperature thermalizes to the local gas temperature.    Typical temperature decrements $(T_{\rm x}(z)-T_{\rm CMB}(z))/(1+z)$
are $\sim$2~K for the 6 cm line and $\sim$1~K for the 1 cm line.  The detection of anti-inverted cm lines will therefore rely critically 
on the filling factor of molecular gas and the line optical depths, both of which will combine to manifest as an effective 
optical depth.  For $\tau_{\rm eff}=0.1$, $\Delta T_{\rm Obs} \simeq 0.1$--0.2~K.   
 
The 100 mas-resolution reference design brightness temperature sensitivity for spectral lines, scaled to 100 km~s$^{-1}$ and 100 hours, 
is 71 K rms at 2.4 GHz (Band 1), 4.4 K at 8 GHz (Band 2), and 0.65 K at 16 GHz (Band 3).  However, this resolution would correspond to 
less than 1 kpc for all redshifts and therefore over-resolve the molecular gas in galaxies (Figure \ref{fig:angsize}).  
In order to beam-match 5 kpc scales at all redshifts, the angular resolution would need to be roughly 600 mas.  At this resolution, 
the rms brightness temperatures are reduced to roughly 2.0 K (Band 1), 122 mK (Band 2), and 18 mK (Band 3).  If one reduces the 
resolution to 1 arcsecond, which will resolve 10 kpc scales at all redshifts, the rms line brightness temperatures for 100 km~s$^{-1}$ 
channels and a 100-hour integration become 0.9 K (Band 1), 55 mK (Band 2), and 8 mK (Band 3).

\section{A ngVLA Formaldehyde Deep Field}\label{sec:fdf}

Given the angular resolution, redshift coverage for single lines and line ratios, and sensitivity considerations above, 
the best compromise ngVLA FDF could be:
\begin{list}{$\bullet$}{\setlength{\topsep}{10pt} \setlength{\itemsep}{0.05in}
		\setlength{\parsep}{0pt} \setlength{\parskip}{0pt}}
 \item  A single 100-hour pointing,
 \item  100 km s$^{-1}$ channels to adequately sample the velocity span of molecular gas in galaxies ($\sim$300~km~s$^{-1}$),
 \item  Full synthesis of Band 2, which will include the 6, 2, and 1 cm H$_2$CO lines 
spanning the molecular history of the universe, $z=0$--7, and
 \item  Angular resolution of 0.6 arcsec, enabling beam-matching to  5 kpc scales at all redshifts.
\end{list}
Line ratios will be available near the peak of the cosmic star formation history, from $z=1.36$ to $z=3.14$.  
Single-line detections will have to be disambiguated using ancillary data such as photometric redshifts, but this
will be straightforward with a carefully-selected FDF location.  While the addition of Band 1 would be highly 
desirable, particularly for density measurements at nearly all redshifts, the brightness temperature sensitivity 
requirements would indicate a prohibitive integration time.

\section{Conclusions}

The ngVLA is uniquely capable of observing a formaldehyde deep field (FDF), which would provide a distance-independent
mass-limited census of molecular gas across the history of star formation and galaxy evolution.  H$_2$CO line ratios in the FDF 
will provide a measurement of the local H$_2$ gas density.  It may also be possible
to make geometric distance measurements over a large
redshift range based on the H$_2$CO $K$-doublet line depths and line ratios \citep{darling2012}.    
A FDF would complement flux-limited ``blind'' molecular emission line surveys \citep[e.g.,][]{pavesi2018}
and could break the usual degeneracy between molecular gas temperature and density encountered in line excitation studies.




\end{document}